\documentclass[12pt]{article}
\usepackage{preamble}
\usepackage{packages}
\usepackage{breqn}
\usepackage{enumerate}
\usepackage[font=small,labelfont=bf]{caption}
\usepackage[left=30mm,right=30mm,top=30mm,bottom=30mm]{geometry}
\usepackage{extarrows}
\usepackage{amsmath,amssymb}
\usepackage{booktabs,subcaption,amsfonts,dcolumn}
\usepackage{braket}
\usepackage{authblk}

\begin{document}
\title{{{\Huge Integrability of Large-Charge Sectors \\ \medskip in Generic 2D EFTs}}}
\author[1,2]{Matthew Dodelson}
\author[2]{Simeon Hellerman}
\author[3,4,5]{Masataka Watanabe}
\author[2,6]{Masahito Yamazaki}
\affil[1]{{\small CERN, Theoretical Physics Department, CH-1211 Geneva 23, Switzerland}}
\affil[2]{{\small Kavli Institute for the Physics and Mathematics of the Universe {\sc (wpi)},\par
The University of Tokyo, Kashiwa, Chiba 277-8582, Japan}}
\affil[3]{{\small Graduate School of Informatics, Nagoya University, Nagoya 464-8601, Japan}}
\affil[4]{{\small Center for Gravitational Physics, Yukawa Institute for Theoretical Physics,\par
Kyoto University, Sakyo-ku, Kyoto 606-8502, Japan}}
\affil[5]{{\small Department of Particle Physics and Astrophysics,\par
Weizmann Institute of Science, Rehovot 7610001, Israel}}
\affil[6]{{\small  Trans-Scale Quantum Science Institute, The University of Tokyo, Tokyo 113-0033, Japan}}
\date{October 2023}
\pagenumbering{gobble}

\maketitle

\begin{abstract}
It is shown that integrability is an accidental property of generic two-dimensional $O(2)$-symmetric asymptotically-free theories in the regime where the charge density is much larger than the dynamical scale. We show this by constructing an infinite tower of higher-spin conserved currents in the most generic effective Lagrangian at large chemical potential to all orders in perturbative expansion in the renormalization-group invariant coupling constant.
\end{abstract}
\newpage
\pagenumbering{arabic}
\pagestyle{plain}


\section{Introduction}

In physics it is notoriously difficult to solve general quantum field theories.
One of the rare exceptions is a special class of theories called integrable field theories in two spacetime dimensions,
theories which have infinitely-many conserved changes and are ``exactly solvable.''
We can extract a large amount of  quantitative information about such theories thanks to  their exact solvability. Moreover, they often serve as useful stepping stones for analyzing more general theories,
and for extracting general physics lessons. One should quickly add, however,
that we expect a generic quantum field theory to be non-integrable.

The goal of this short note is to point out the existence of classical integrability for a large class of 
effective field theories (EFTs) in two dimensions. Our EFTs are {\it generic}, the
only requirement being that the theory arises from the large-charge sector \cite{Hellerman:2015nra,Alvarez-Gaume:2016vff,Monin:2016jmo,Hellerman:2017efx,Jafferis:2017zna,Hellerman:2018sjf,Watanabe:2019pdh,Badel:2019oxl,Antipin:2020abu,Antipin:2020rdw,Gaume:2020bmp} of an $O(2)$-symmetric asymptotically free theory.\footnote{This paper extends the analysis of the large-charge sectors, previously done in three and higher dimensions, to two spacetime dimensions.}
Since our discussion is within EFT,
our analysis automatically incorporates many examples of EFT arising via Renormalization-Group (RG) flows from 
well-defined UV theories, such as non-linear sigma models.

The rest of this paper is organized as follows.
After a summary of the large-charge EFT with $O(2)$ symmetry  in section \ref{sec:LargeCharge},
we present a proof of existence of higher-spin conserved currents in section \ref{sec:ConservedCurrent}.
We end in section \ref{sec:Discussion} with comments and discussions.

\section{Large-charge EFT for asymptotically free theories} \label{sec:LargeCharge}

Let us consider an asymptotically-free QFT  $\mathcal{T}_{\rm UV}$
with an $O(2)$ global symmetry in $D$ spacetime dimensions.
(We will restrict to $D=2$ in the next section.)
This theory can be regarded as the ``UV theory'' in our discussion.
Note that we do {\it not} need to assume that this theory is integrable.

Since the theory is asymptotically-free, we expect the theory to generate a dynamical scale, which we denote by $M$. For our considerations we also introduce an IR regulator $1/R$, where $R$ can be regarded as the size of the spatial
sphere $S^{D-1}$. We need a  hierarchy $M \gg 1/R$. In general these are the only mass scales
we expect in the theory. In general we expect the theory to be strongly coupled in the IR,
where we do not expect perturbative techniques to be applicable.

Let us now consider the setup with a very large $O(2)$ charge $Q \gg 1$. The crucial difference from our previous discussion 
is that we now have a new mass scale $\mu$, as determined by the charge density $\rho$ on the sphere:
$\mu^{D-1} \sim \rho \sim Q/ R^{D-1}$. In practice we can regard this scale as the chemical potential for the $O(2)$ symmetry,
i.e.\ the Vacuum Expectation Value (VEV) $\langle A_0 \rangle$ of the 
time component of the background $O(2)$ gauge field $A_{\mu}$. 

Let us consider the situation where $Q$ is very large such that $\mu$ is above the dynamical scale:
\begin{align}
\mu \gg M  \gg 1/R \;.
\end{align}

Due to the introduction of the chemical potential, the fields charged under the $O(2)$ symmetry
will
have a mass of $O(\mu)$, and hence can be integrated out below the scale $\mu$. In particular, 
we expect that the running of the coupling constant stops at the energy scale $\mu$, and 
when $\mu\gg M$ we expect the theory to have a weakly-coupled EFT description. We call this theory $\mathcal{T}_{\rm EFT}$.\footnote{In the literature one often takes the small radius  limit $R\to 0$, to create the hierarchy $1/R\gg M$, to appeal to weakly-coupled descriptions. One might then hope to extrapolate back to the large-radius limit $R\to \infty$. Here the radius $R$ is kept large, and we instead generate a new scale $\mu$ by 
focusing on the large-charge sector.} 

Which degrees of freedom should be kept in the EFT? A natural candidate is obtained by considering the complex
``order parameter'' $\Phi = |\Phi| e^{i\theta}$: its phase is shifted by the $O(2)$ symmetry  transformation
$A_{\mu}\to A_{\mu} + \partial_{\mu} \chi$ 
as $\theta\to \theta + \chi$, and $\theta$ represents the flat direction respecting the $O(2)$ symmetry.

Since we are introducing the chemical potential, we have 
the shift $A_0\to A_0 + \mu$, which can be traded for a shift $\theta\to \theta + \mu t$.
This means that we need to re-expand the theory around a new vacuum as
\begin{align}
\theta=\mu t + \theta' \;,
\label{eq:theta}
\end{align}
where $\theta'$ represents a fluctuation around the new vacuum.
The newly-chosen vacuum breaks the $O(2)$ symmetry, and 
$\theta$ can be light degrees of freedom (Nambu-Goldstone mode) of the broken $O(2)$ symmetry.

For the compatibility with the $O(2)$ shift symmetry,
we only expect derivatives of $\theta$ to appear in the Lagrangian.
Moreover, terms with higher derivatives will be suppressed by the 
scale separation of the theory, which is $1/\mu$ in this case.
Finally, the Lagrangian should be compatible with the underlying Lorentz symmetry.
We can thus write down the Lagrangian for the EFT $\mathcal{T}_{\rm EFT}$ to
be the ``free'' (indeed free when $D=2$) Lagrangian, up to corrections suppressed by $\mu^{-1} \propto Q^{-1/(D-1)}$:
\begin{align}
    \mathcal{L} = \frac{1}{g_{\rm EFT}^2}\abs{\partial \theta}^D+O\left(Q^{-\frac{1}{D-1}}\right) = \frac{2}{g_{\rm EFT}^2} K +O\left(Q^{-\frac{1}{D-1}}\right)\;,
    \quad K\equiv \frac{1}{2}\abs{\partial \theta}^D \;,
\end{align}
where $\abs{\partial \theta}\equiv \sqrt{\partial_\mu\theta\partial^\mu \theta}$ and $\theta$ is to be expanded as in \eqref{eq:theta}.

The EFT coupling constant $g_{\rm EFT}$ does not run under the RG  flow inside the EFT, since we have already integrated out
all the degrees of freedom other than $\theta$. 
Its value can be matched
 with the gauge coupling $g_{\rm UV}$ of the UV theory---for example,
with the coupling constant of the UV $O(N)$ theory.
Even though our theories are weakly-coupled, $g_{\rm UV}$ has RG running,
which can be accounted by the one-loop and higher-loop computations, and we can match the value of $g_{\rm EFT}$ with that of $g_{\rm UV}$ RG-evolved down to the energy scale $\mu$.
Since the details will depend on the choice of the UV theory, 
let us here express this as
\begin{align}
    \mathcal{L} = f(\ell)K+ \text{(subleading in $1/Q$)},
    \quad \text{where} \quad \ell\equiv \log\left(\frac{2K}{M^2}\right),
\end{align}
where the coupling constant has been replaced by a function $f(\ell)$.
We can 
write this more concisely as 
\begin{align}
    \mathcal{L} = \mathcal{F}(K)+ \text{(subleading in $1/Q$)} \;.
\end{align}
where we make the dependence on the dynamical scale implicit.

\section{Higher-spin currents in generic large-charge EFT}\label{sec:ConservedCurrent}

\subsection{Construction of the higher-spin currents}

In the rest of this paper we specialize to $D=2$ spacetime dimensions. We will work with the EFT Lagrangian at (exactly) \textit{leading order} in the large charge expansion,
\begin{align}
\mathcal{L} = \mathcal{F}(K) \;.
\label{eq:most-generic-EFT}
\end{align}
This corresponds to taking all the possibilities for the perturbative corrections while ignoring all non-perturbative corrections in $g_{\rm UV}$.
We will use the light-cone coordinates $ds^2 = 2 dx^+ dx^-$,
with metric $g_{+-}=g^{+-}=1, g_{++}=g_{--}=g^{++}=g^{--}=0$.
This means, for example, $K = \frac{1}{2} (\partial \theta)^2  = (\partial_+ \theta) (\partial_- \theta)$.
The equation of motion of this theory is given by
\begin{align}
\label{eq:eom}
\theta_{,+-} = \mathcal{G}(K) \left[
   \frac{\theta_{,-}}{\theta_{,+}}  \theta_{,++}  + \frac{\theta_{,+}}{\theta_{,-}}  \theta_{,--} 
\right] \;,
\end{align}
where we defined
\begin{align}
\mathcal{G}(K) \equiv -\frac{1}{2} \frac{K \mathcal{F}''(K)}{\mathcal{F}'(K) + K \mathcal{F}''(K)} \;.
\end{align}

We claim that the most generic form of the effective Lagrangian \eqref{eq:most-generic-EFT} allows for an infinite tower of higher-spin conserved currents:
for any positive integer $k$ we find a spin-$2k$ current, which is conserved on-shell.\footnote{While these currents exist as formal objects for any complex value of $k$, single-valuedness of correlation functions in the EFT restricts $k$ to be integer or half-integer.  More generally objects with arbitrary fractional powers of the product $\partial_+\theta \partial_-\theta$ exist in the EFT but generically do not come from operators in the UV theory.  }
Since these currents have different spins it immediately follows that these charges are linearly independent,
and we have an infinite family of independent conserved charges, making the EFT integrable.\footnote{More precisely we find integrability in the massless sector of the EFT. It is an interesting question to extend the discussion to massive Goldstone sectors, see e.g.\ \cite{Watanabe:2013uya,Cuomo:2020gyl} for related discussion.}
Since this is a statement about the EFT,
this holds irrespective of the choice of the UV theory.

Our discussion of integrability uses the equation of motion, and is purely classical.
However, it is also the case that our analysis applies to a general EFT,
and we can thus incorporate quantum effects from RG flows to the 
parameters in the classical Lagrangian.

We are interested in constructing a spin-$2k$ conserved current 
with components $(T_{2k}, \Theta_{2k-2})$, 
satisfying the conservation condition
\begin{align}
    \partial_{+} T_{2k} = \partial_{-}\Theta_{2k-2} \;.
    \label{eq:conservation}
\end{align}
This will lead to spin-$(2k-1)$ conserved charges\footnote{In our analysis integer and half-integer values of $k$ appear on equal footing.}
\begin{align}
Q_{2k-1} = \int dx^1\, \left(T_{2k}
 + \Theta_{2k-2} \right) \;.
\end{align}

For $k=1$ we have a spin-$2$ current, which we identify with the stress-energy tensor as expected.
In the following we will consider the case $k>1$, since these are the charges relevant for integrability.

The components $T_{2k}$ and $\Theta_{2k-2}$ of the current can be expressed as
\begin{align}
&T_{2k}= A_{2k}(\ell) (\partial_{-} \theta)^{2k} \;,\\
&\Theta_{2k-2}= B_{2k}(\ell) (\partial_{+} \theta) (\partial_{-} \theta)^{2k-1} \;.
\end{align}
for $\ell$-dependent functions 
$A_{2k}(\ell), B_{2k}(\ell)$.
Note that this form is by no means special (for example, we have discarded terms including higher derivatives of $\theta$). We will however see that this special form is enough to construct one set of higher-spin currents by explicitly finding a candidate for $A$ and $B$ which exactly satisfies the conservation law.
For $k>1$, $A_{2k}(\ell)$ is a solution to a second-order ODE
\begin{align}\label{eq:ODE_A}
\frac{A''_{2k}(\ell)}{h(\ell)}
+ \left[\frac{2k-1}{h(\ell)} -1 -\frac{h'(\ell)}{h^2(\ell)} \right]
A'_{2k}(\ell) - k(2k-1) A_{2k}(\ell) =0 \;,
\end{align}
and $B_{2k}(\ell)$ is determined from $A_{2k}(\ell)$ as 
\begin{align}
B_{2k}(\ell) = \frac{1}{k-1} \left[-k A_{2k}(\ell) + \frac{A'_{2k}(\ell)}{h(\ell)} \right] \;,
\end{align}
where we defined
\begin{align}
&h(\ell) \equiv \frac{d \log \mathcal{F}'(K)}{d\ell} 
= \frac{K\mathcal{F}''(K)}{\mathcal{F}'(K)}
=-\frac{2\mathcal{G}(K)}{2\mathcal{G}(K)+1} \;.
\end{align}
This means that a tower of higher-spin currents exists whenever there is a solution to $A_{2k}$ for all values of $k>1$.

\subsection{Proof of conservation equations}

Let us show the current conservation by explicit computations.
We compute
\begin{align}
\partial_{+} T_{2k} &= [2k A_{2k}(\ell) + A'_{2k}(\ell) ] (\theta_{,-})^{2k-1} (\theta_{,+-})  + A'_{2k}(\ell) (\theta_{,-})^{2k} \frac{\theta_{,++}}{\theta_{,+}} \;,\\
\partial_{-} \Theta_{2k-2} &= [(2k-1) B_{2k}(\ell) + B'_{2k}(\ell) ] (\theta_{,-})^{2k-2} (\theta_{,+})  (\theta_{,--}) \nonumber\\
& \qquad \qquad +[B_{2k}(\ell) +  B'_{2k}(\ell)] (\theta_{,-})^{2k-1} (\theta_{,+-})\;,
\end{align}
and we obtain the conservation equation \eqref{eq:conservation} by equating the two expressions.

Thanks to the equation of motion \eqref{eq:eom},
we can eliminate $\theta_{,+-}$ in terms of $\theta_{,++}$ and $\theta_{,--}$,
to convert the conservation equations into two linear first-order ordinary differential equations:
\begin{align}
&A'_{2k}(\ell)  = -(2k-1) B_{2k}(\ell) 
- B'_{2k}(\ell) \;,
\label{eq:Ap_B_Bp}\\
&\mathcal{G}(K) [2k A_{2k}(\ell) + A'_{2k}(\ell) ]
 + A'_{2k}(\ell) = \mathcal{G}(K) [B_{2k}(\ell) + B'_{2k}(\ell)] \;.
\end{align}
Note that the first (second) equation is independent of (depends on) the function $\mathcal{F}[K]$, and hence of the choice of the EFT.

From these equations we can eliminate the functions $B_{2k}(\ell), B'_{2k}(\ell)$ as 
\begin{align}
& B_{2k}(\ell) = \frac{1}{k-1} \left[-k A_{2k}(\ell) + \frac{A'_{2k}(\ell)}{h(\ell)}\right] \;, \\
& B'_{2k}(\ell) = \frac{1}{k-1} \left[-\left(k + \frac{h'(\ell)}{h^2(\ell)} \right)  A'_{2k}(\ell) + \frac{A''_{2k}(\ell)}{h(\ell)}\right] \;.
\end{align}
    After plugging these expressions back into \eqref{eq:Ap_B_Bp},
we obtain the ODE for $A_{2k}(\ell)$ as in 
\eqref{eq:ODE_A}.

\subsection{Example}

While the existence of solutions to the ODE
\eqref{eq:ODE_A} is sufficient for the integrability of the EFT,
one might be tempted to find analytic expressions for the conserved current.
This is possible for the ``leading-log'' EFT, for which the Lagrangian takes the form
\begin{align}
\mathcal{L} = \mathcal{F}[K]= \beta \left(\log \frac{2K}{M^2} -1 \right) K \;,
\end{align}
which leads to $\mathcal{F}'[K] = \beta \ell$ and $h[\ell] = 1/\ell$.

In this case, the ODE for $A_{2k}(\ell)$ reads 
\begin{align}
\ell A''_{2k}(\ell) + (2k-1) \ell A'_{2k}(\ell) - k(2k-1) A_{2k}(\ell) = 0 \;.
\end{align}
One of the solution to this equation is
\begin{align}
A_{2k}(\ell) = L_k^{(-1)} [ -(2k-1) \ell ]\;,
\end{align}
where $L_k^{(\alpha)}[x]$ is a generalized Laguerre polynomial of order $k$ and argument $x$ (for $\alpha=0$ this reduces to the ordinary Laguerre polynomial).\footnote{Since the ODE is second order, there are two independent solutions. Here we have shown solutions that are polynomial in $\ell$. The other is given by the Meijer's $G$-function $G_{1\,2}^{2\, 0}\left(-(2k-1)\ell\left|\begin{matrix}
    1+k\\
    0,\, 1
\end{matrix}\right.\right)$, which is expanded as $\textstyle e^{-(2k-1)\ell}\sum_i 1/\ell^i$.}
For example, we have
\begin{align}
\begin{split}
A_4(\ell)& = \ell (3\ell+2)\;, \\
A_6(\ell)& = \ell (25\ell^2 + 30 \ell+6) \;,\\
A_8(\ell)& = \ell (343\ell^3 + 588\ell^2+252\ell+24) \;.
\end{split}
\end{align}

\section{Discussions}\label{sec:Discussion}


In the standard discussion of integrability, it is often stated that classical integrability implies
factorization of the tree-level S-matrix, at least in 
 the integrable subsector. In particular, the theory allows for 
no particle production, and this severely constrains the form of the Lagrangian, 
at least order by order in perturbation theory (see \cite{Shankar:1977cm,Zamolodchikov:1978xm,Parke:1980ki,Dorey:1996gd}).
This is in apparent contradiction with our statement that
integrability exists for a generic large-charge EFT.

One should note, however, that the connection between classical integrability and factorization of the tree-level S-matrix
is in general broken for  integrable two-dimensional field theories with massless excitations \cite{Hoare:2018jim}:
our EFT applies to a massless mode $\theta$ associated with the symmetry breaking. 
Related to this, our theory is an EFT and its kinetic term is 
non-canonical, contrary to what is often assumed in the literature. In short, we do not in general
expect factorization of the tree-level S-matrix in our theories. In fact, it is not obvious whether the S-matrix in these theories is well-defined, due to the presence of IR divergences.\footnote{Here we are referring to the S-matrix of fluctuations around the large charge vacuum. The S-matrix around the trivial vacuum $\theta=0$ is IR finite (see e.g. \cite{Dubovsky:2012sh}).  However, the higher spin charges act nonlinearly around the trivial vacuum, so the usual argument \cite{Parke:1980ki,Dorey:1996gd} for the factorization of the S-matrix does not apply. We thank Sergei Dubovsky for raising this question.} It would be interesting to understand whether one can construct an S-matrix for the EFT, or if there is some other dynamical implication of the infinite family of charges.

There are several directions for future research. First, it is natural to extend our analysis to 
large-charge EFTs associated with more general theories, such as EFTs with $O(N)$ symmetries with $N\ge 2$.
Second, it would be interesting to compare our analysis with 
detailed analysis of specific UV integrable field theories, 
such as the $O(N)$ model \cite{DiPietro:2021yxb,Marino:2021dzn}.
It would also be interesting to see how the integrability of EFT fits into the framework of four-dimensional Chern-Simons theory \cite{Costello:2019tri,Ashwinkumar:2023zbu}.

\section*{Acknowledgements}

We would like to thank Arkady Tseytlin and Heng-Yu Chen for discussions.
This work is supported in part by WPI Research Center Initiative, MEXT,
Japan. 
This work was also supported by Israel Science Foundation center for excellence grant (grant number 2289/18) and by the German Research Foundation through a German-Israeli Project Cooperation (DIP) grant ``Holography and the Swampland''.
MD is supported by the European Research Council (ERC) under the European Union’s Horizon 2020 research and innovation programme (grant agreement number 949077). 
MD was also supported by JSPS KAKENHI Grant Number 20K14465.
The work of SH was supported in part by JSPS KAKENHI
Grant Numbers JP22740153 and JP26400242.  SH also thanks the Simons
Center for Geometry and Physics and IHES, for hospitality while this work was in progress.
MW is supported by Grant-in-Aid for 
JSPS Fellows (No.\ 22J00752).
MW was also supported by the Foreign Postdoctoral Fellowship Program of the Israel Academy of Sciences and Humanities.
MW thanks the hospitality of Kavli IPMU while this work was in progress.
MY is supported by the JSPS Grant-in-Aid for Scientific Research
19H00689, 19K03820, 20H05860, 23H01168, and 23K17689, and by
JST Presto Grant Number  JPMJPR225A and by JST [Moonshot R\&D] Grant No.\ JPMJMS2061. 

\bibliographystyle{JHEP}
\bibliography{ref}
\end{document}